\documentstyle[aps,psfig,floats,twocolumn]{revtex}

\begin{document}

\twocolumn[
\hsize\textwidth\columnwidth\hsize\csname @twocolumnfalse\endcsname

\title{Impact of Electron-Electron Cusp on Configuration Interaction Energies}
\author{David Prendergast,$^{a)}$ M. Nolan,$^{b)}$ Claudia Filippi,$^{a)}$ 
Stephen Fahy,$^{a)}$ and J.C. Greer$^{b)}$}
\address{$^{a)}$Department of Physics and $^{b)}$NMRC, University College, Cork, Ireland}
\date{\today}
\maketitle


\begin{abstract}
The effect of the electron-electron cusp on the convergence of 
configuration interaction (CI) wave functions is examined. 
By analogy with the pseudopotential approach for electron-ion interactions, 
an effective electron-electron interaction is developed which closely 
reproduces the scattering of the Coulomb interaction but is smooth and 
finite at zero electron-electron separation. 
The exact many-electron wave function for this smooth effective interaction 
has no cusp at zero electron-electron separation. 
We perform CI and quantum Monte Carlo calculations for He and Be atoms, 
both with the Coulomb electron-electron interaction and with the smooth 
effective electron-electron interaction.  
We find that convergence of the CI expansion of the wave function for the 
smooth electron-electron interaction is not significantly improved compared 
with that for the divergent Coulomb interaction for energy differences on 
the order of 1 mHartree. 
This shows that, contrary to popular belief, description of the 
electron-electron cusp is not a limiting factor, to within
chemical accuracy, for CI calculations.
  
\end{abstract}

]

\section{Introduction}
%
The primary difficulty in solving the Schr\"odinger equation for
many-electron systems arises from the presence of the electron-electron 
interaction, which results in a problem non-separable in the electron
coordinates.
Configuration interaction (CI) calculations use an expansion of the
many-body wave function in configuration state functions (CSF) 
given by spin coupled sums of determinants built from single-particle 
orbitals. 
One of the central difficulties for the convergence of the CI 
expansion is thought to arise from the behaviour of the Coulomb 
interaction at small electron-electron separations \cite{%
KlahnMorg,Hill,Morgan:CIconv,r12,KutzMorg,Hoff,Helgaker,%
Lein,NooBar,Silanes}.
The many-body wave function must have an appropriate cusp at 
electron coalescences so that the infinite Coulomb 
interaction term is exactly cancelled by an opposite divergence 
of the kinetic energy~\cite{Kato:cusp}.

It is often stated in the literature that failure of the CI expansion
to reproduce the correct electron cusp, i.e. the short-range part of 
the Coulomb hole, leads to a slow convergence in the energy with 
respect to the number of CSF's~\cite{Morgan:CIconv}.
While this has been shown~\cite{Hill,KutzMorg} in the asymptotic regime 
(i.e.\ very large angular momentum $l$ and energy errors of the order 
of $10^{-6}$ Hartree), it is not clear that it applies in the 
practical regime (mHartree accuracy) for multi-electron systems.

Explicit inclusion of terms in the interelectronic distance 
(r$_{12}$ methods)~\cite{Hyll,r12}
or the use of correlated (geminal) Gaussian basis functions~\cite{gem_gauss} 
significantly improve the convergence of a CI calculation.
For the helium atom, the inclusion of r$_{12}$ terms~\cite{KutzMorg} 
yields an expansion in partial waves converging asymtotically as 
$l^{-6}$ or better, instead of a slow $l^{-3}$,
where $l$ is the maximum angular momentum in the expansion.
Baker {\it et al.}~\cite{BFHM90} also showed that using basis 
functions with the same analytic structure as the true wave 
function (depending on powers and logarithms of $r_{12}$) 
leads to a reduced expansion length. 
The success of methods explicitly including $r_{12}$ terms has been 
attributed to their correct description of the short-range part 
of the interaction.

A completely different point of view, taken by Gilbert in 1963~\cite{Gil63},
is that the difficulty of CI wave functions in describing short-range 
effects of the cusp are energetically unimportant and that the 
benefit of the explicitly correlated approach is that the {\em entire} 
Coulomb hole, with a size on the order of the atomic orbitals, is 
much easier to describe in terms of interelectronic coordinates.
Gilbert's suggestion appears to have received little attention and the 
idea that smoothing out the Coulomb potential and {\em removing the cusp 
altogether} will automatically lead to an improved CI convergence has 
been stated in the literature \cite{King,NooBar}.
However, no systematic studies of the effect of the electron cusp 
on CI energy convergence at the level of interest for chemical 
accuracy (1 mHartree for total energies) have been presented.

In this paper, we explore whether a significant improvement in the 
convergence of the CI expansion is obtained when the true Hamiltonian 
is replaced by one which has a smooth and finite electron-electron 
interaction (pseudo-interaction) at short distances and, 
consequently, smooth eigenfunctions with no cusps.

Surprisingly, in calculations for the He and Be atoms, we
find that removing the divergence of the Coulomb potential
at electron-electron coalescences does not significantly reduce 
the length of the CI expansion, thereby substantiating the argument 
put forward by Gilbert~\cite{Gil63} that the slow convergence should 
be attributed to difficulties in describing, not the interelectron 
cusp, but the intermediate-range electronic correlation.
The poor treatment of the interelectron cusp in the CI expansion
is not a limiting factor at the mHartree level of accuracy on 
the total energy.
To further confirm this result, we calculate the energy within 
quantum Monte Carlo (QMC) for the true electron-electron 
interaction with accurate wave functions which explicitly 
contain interelectronic coordinates, and then deliberately remove the 
short-range correlation term related to the cusp, keeping other parts 
of the wave function unchanged. We find that the removal of the cusp
has a very small effect on the variational energy.
Finally, we demonstrate that the convergence behavior of the CI expansion
as a function of angular momentum in the range practical for many 
applications ($l<6$) is largely unaffected by the smoothing of the 
electron cusps.
Therefore, the angular momentum convergence numerically observed 
by other authors~\cite{Helgaker} is related to problems associated 
with intermediate range correlations, not the cusp region as such.

In Sec.~\ref{methods}, we describe the pseudopotential used for
the electron-electron interaction and briefly present the QMC
and CI methods.
In Sec.~\ref{results}, we report the QMC and CI energies for He 
and Be atoms using the true and pseudo electron interaction.
In Appendices~\ref{app:gen_psint} and \ref{app:acc_psint}, we discuss the
numerical generation and accuracy of the electron pseudo-interaction, 
and in Appendix~\ref{app:srjast} we define a short-range correlation
term which can be associated with it.
In Appendix~\ref{app:two_cent_int}, we briefly discuss the
evaluation of two-electron integrals for the pseudo-interaction.

\section{Computational Methods}
\label{methods}
%
%
\subsection{Electron-electron Pseudo-interaction}
\label{methods:pseudoint}

When modifying the short-distance electron-electron interaction,
we wish to keep the intermediate- and long-range correlation 
properties of the many-body eigenstates unchanged.
This will allow us to isolate systematically the effects of the short-range 
cusp-like behaviour in the wave function.
To achieve this, our new (smooth) interaction should ideally 
have electron-electron scattering properties identical to the true 
interaction for all energies.
If this were the case, all many-electron eigenvalues would be identical 
for both interactions.
While this ideal cannot be exactly realized, we expect that a potential 
with the same scattering properties as the true interaction over 
a sufficiently wide range of scattering energies will give many-electron 
eigenvalues very close to the true ones. 

This goal for electron-electron scattering is analogous to the 
problem of electron-ion scattering addressed in the well-established 
pseudopotential method~\cite{HSC}.
In that approach, the aim is to produce a smooth electron-ion potential 
which reproduces the single-particle eigenvalues of the true system 
over a range of energies, and has the same single-particle wave 
functions outside some suitably chosen cut-off radius $r_c$.
Such pseudopotentials are generated for the isolated atom but may 
then be used with a high level of accuracy in molecules and solids.
In solids, the smoothness of the potential greatly improves the 
convergence of plane-wave expansions for single-particle orbitals
\cite{Pickett}.

Following the success of norm-conserving pseudopotential methods 
in the context of the electron-ion scattering problem, we generate 
a norm-conserving pseudo-interaction for the two-electron scattering 
problem, which will replace the true 
$e^2/r$ electron-electron interaction inside a cut-off radius $r_c$.
Since the electron-electron interaction is repulsive and has no bound 
states for an isolated pair of electrons, we use, with some minor
modifications, the generalized norm-conserving method devised by 
Hamann~\cite{Hamann:gncpp} to generate pseudopotentials for
electron-ion interactions at unbound, scattering state energies.
The numerical details of this approach are presented in 
Appendix~\ref{app:gen_psint}.

In Fig.~\ref{fig:interxns}, we show a pseudo-interaction potential
$V_{\rm ps}$ generated by this approach, along with the true $e^2/r$ 
interaction and a smooth but not norm-conserving interaction, 
$V_{\rm erf}(r) = e^2{\rm erf}(r/r_c)/r$,
as used by Savin and coworkers~\cite{Savin:erf} in a different
context.
In Appendix~\ref{app:acc_psint}, the accuracy of such scattering 
potentials in reproducing the exact eigenstate energies is analyzed 
qualitatively using first-order perturbation theory. 
In Appendix~\ref{app:srjast}, we define a short-range inter-electronic 
correlation term, which can be directly related to the replacement 
of the divergent Coulomb interaction with a smooth norm-conserving 
interaction. 

%
%

\begin{figure}[!tb]
\begin{center}
  \leavevmode
  \vspace{1.0cm}
  \psfig{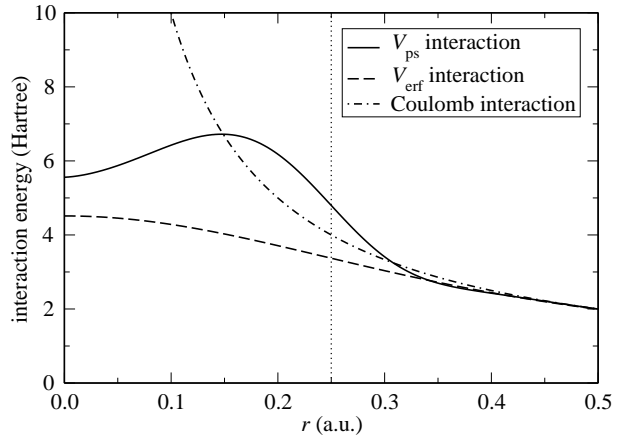}
  \vspace{0.5cm}
  \caption{
    Norm-conserving electron pseudo-interaction (solid line)
    and smooth interaction $V_{\rm erf}$ (dashed line),
    each with a cut-off radius $r_c = 0.25$ a.u. (indicated by a vertical dotted line),
    and the true $1/r$ interaction (dash-dot line),
    as functions of the electron separation $r$.
  }
  \label{fig:interxns}
\end{center}
\end{figure}

\subsection{Quantum Monte Carlo Calculations}

We employ QMC methods to determine accurate energies 
both for the Coulomb and the pseudo-interaction.
The wave functions $\Psi$ used in QMC are of the form given 
in Ref.~\cite{Fil-Umr} and explicitly depend on the 
inter-electronic coordinates. 
They include parameterized terms to describe electron-nucleus, 
electron-electron and electron-electron-nucleus correlation, 
as follows:
\begin{eqnarray}
\Psi=\left(\sum_n d_n {\rm D}^\uparrow_n {\rm D}^\downarrow_n\right)
{\rm J}(r_i, r_j, r_{ij}) ~. 
\label{eq:jast_slt}
\end{eqnarray}
${\rm D}^\uparrow_n$ and ${\rm D}^\downarrow_n$ are the Slater 
determinants of single particle orbitals for the spin-up and down 
electrons, respectively, and $J$ is a Jastrow factor of the form,
\begin{eqnarray}
{\rm J}(r_i,r_j,r_{ij})=\prod_{i} \;\exp{(A_{i})}
\;\prod_{ij}\;\exp{(B_{ij})} \;\prod_{ij}\;\exp{(P_{ij})},
\end{eqnarray}
where $A$ is an electron-nuclear term (which could be omitted if
a sufficiently large single particle basis were used in constructing
the determinantal part), $B$ is an electron-electron term 
incorporating the cusp at small $r_{ij}$, and $P$ is a smooth
function of the electron coordinates, $r_i$ and $r_j$, as 
discussed in detail in Ref.~\cite{Fil-Umr}.
For the pseudo-interaction calculations, the $B$ term is omitted.
The parameters in the determinant and Jastrow factor are optimized 
within QMC using the variance minimisation method~\cite{Cyrus:varmin} 
and the accuracy is tested at the variational level.
The wave function is then used in diffusion Monte Carlo (DMC), 
which produces the best energy within the fixed-node approximation 
(i.e.\ the lowest-energy state with the same nodes as the trial wave 
function).

For a given interaction, this method yields the exact eigenvalue (within 
statistical sampling noise) for the ground state of the He isoelectronic 
series since the singlet ground state wave function of a two-electron system 
is nodeless.
For He, only one determinant is used in the wave function.
For the Be atom,  the 2s and 2p configurations are included in
the determinantal part of the wave function and DMC gives the best 
energy subject to the fixed-node condition. 
To isolate the effect of the pseudo-interaction from the fixed-node 
approximation, we keep the nodal structure of the wave 
function fixed while modifying the electron-electron interaction.

\subsection{Configuration Interaction Calculations}

The CI calculations are performed using the Monte Carlo configuration 
interaction (MCCI) method~\cite{Gre1,Gre2}. 
In a MCCI calculation, a coefficient threshold $c_{\rm min}$ is defined, 
configurations are randomly generated, and, after a variational 
calculation, coefficients of magnitude below 
$c_{\rm min}$ are rejected. 
This procedure is repeated until a desired degree of convergence in the 
CI calculation is achieved. 
A full CI (FCI) calculation can be obtained for $c_{\rm min}= 0$. 
The fact that the MCCI procedure does not rely on a pre-selection of 
CSF's makes it well-suited for our study: the dynamical selection of 
configurations allows for rejection of CSF's which may no longer be
needed with the smooth electron interaction.

For use in the CI approach, the pseudo-interaction $V_{\rm ps}$ is 
represented (in Hartree atomic units) as a sum of the smooth function 
$V_{\rm erf}$ plus $n_G$ Gaussian terms, as follows:
\begin{equation}
V_{\rm ps}(r) ~=~ \frac{{\rm erf}(r/r_c)}{r} + \sum_{i = 1}^{n_G} c_i\, 
{\rm exp}(-d_i r^2) ~.
\label{eq:psfit}
\end{equation}
This reproduces the $1/r$ tail of the Coulomb interaction and the
smooth, finite form of the pseudo-interaction at short distances.
The parameters $c_i$ and $d_i$ are found by least-squares fitting 
(for radii $r<3r_c$) to the numerical potential, generated as 
described in Appendix~\ref{app:gen_psint}.
For the CI results presented below, the potential was fitted 
using $n_G=9$.
QMC calculations of the He atom energy with the numerical potential 
and with the fitted potential show a negligible difference (less than
$0.02$ mHartree for all values of $r_c<1$ a.u.).

Evaluation of the pseudo-interaction $V_{\rm ps}(r)$ can be 
implemented in codes for Gaussian integral calculation since it 
is possible to compute the two-body matrix elements both of 
$\exp(-d\; r^2)$ and ${\rm erf}(r/r_c)/r$ in the CI basis.
Matrix elements of the interaction ${\rm erf}(r/r_c)/r$ for all
s-type basis functions have already been given in Ref.~\cite{Savin:erf}.
For both types of interaction, a simple modification 
to the integral codes is possible if the standard two-electron 
integrals are calculated using the method of Rys quadrature~\cite{drk}.
Details of two-body integral evaluation with the modified form of the 
electron interaction are given in Appendix~\ref{app:two_cent_int}.

%
%
\section{H\lowercase{e} and B\lowercase{e} atomic energies with pseudo-interactions}
\label{results}
%
%

To test the impact of the interelectron cusp on the energy, 
we calculate the total energies of He and Be atoms, replacing 
the true $1/r$ electron interaction with the norm-conserving 
pseudo-interaction $V_{\rm ps}(r)$, generated for various values 
of the cut-off radius $r_c$.
For comparison, we also present results for the smooth but not 
norm-conserving potential $V_{\rm erf}(r) = {\rm erf}(r/r_c)/r$. 
The results in this section are obtained from QMC and CI calculations. 

%
%
\subsection{Ground State of the He atom}
%
%

\begin{figure}[tb]
\begin{center}
  \leavevmode
  \psfig{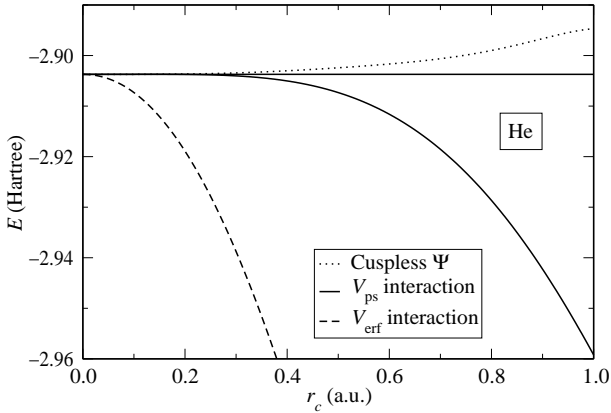}
  \vspace{0.5cm}
  \caption{
    He atom ground state energy $E$ as a function of cut-off radius $r_c$, 
    obtained from DMC calculations using various 
    forms of the electron-electron interaction:
    (a) the norm-conserving pseudo-interaction $V_{\rm ps}$ 
        (solid line);
    (b) the smooth interaction $V_{\rm erf}$ 
        (dashed line).
    The curves are polynomial fits to the calculated DMC values, 
    as discussed in the text.
    The horizontal line shows the exact value~\protect\cite{Davidson} for the He ground state.
    Also shown are variational energies, with the true Coulomb interaction,
    upon removal of the electron cusp (see Appendix~\protect\ref{app:srjast})
    from the wave function used in the DMC calculations (dotted line).
  }
  \label{fig:HegsE}
\end{center}
\end{figure}

In Fig.~\ref{fig:HegsE}, we show
the singlet ground state energies of the He atom calculated 
with a norm-conserving electron-electron interaction $V_{\rm ps}$ 
generated as in Sec.~\ref{methods}, for various values of the 
cut-off radius $r_c$.
The values obtained from QMC (exact) calculations are shown.
%
%
We also plot the energies calculated using the $V_{\rm erf}(r)$ interaction.
The horizontal line indicates the exact energy $E_0 = -2.903724$ Hartree 
for the true Coulomb interaction~\cite{Davidson}. 

%
%

The calculated QMC energies for $V_{\rm ps}$ and $V_{\rm erf}$ 
in Fig.~\ref{fig:HegsE}
are fitted with polynomial expansions of the form:
$E(r_c) = E_0 + a r_c^2 + b r_c^3 + c r_c^4 + d r_c^5 + e r_c^6$.
As discussed in Appendix~\ref{app:acc_psint}, any potential which 
differs from the exact $1/r$ interaction would be expected, from 
first order perturbation theory, to converge at least quadratically 
in $r_c$ to the exact energy $E_0$ for small $r_c$. 
Since $V_{\rm erf}$ is not explicitly 
constructed to eliminate first order errors in $V_{\rm erf}(r)-1/r$,
quadratic convergence is seen for $V_{\rm erf}$ and
the fitting parameters are: 
$a=-0.3279$, $b=-0.3839$, $c=0.6758$, $d=-0.2622$, and $e=0$.
On the other hand, the norm-conserving pseudo-interaction 
construction of Sec.~\ref{methods:pseudoint} explicitly ensures that 
the first-order perturbation correction to the energy is zero.
The He ground state energy as a function of $r_c$ using $V_{\rm ps}$ 
can indeed be fitted without the quadratic term, and even the cubic term 
proves to be small,
the polynomial coefficients being:
$a=0$, $b=0.0188$, $c=-0.1023$, $d=0$, and $e=0.0280$.
For He, the energies obtained using $V_{\rm ps}(r)$ are within 1 mHartree 
of the energy computed with the true Coulomb interaction for values
of $r_c < 0.4$ a.u. 
Since the typical electron-electron distance in the He ground state 
is of the order of 1 a.u., these results indicate that the first peak 
of the electron pair-distribution function provides a good measure 
of how large a cut-off radius $r_c$ may be used to obtain accurate 
total energies with $V_{\rm ps}(r)$: 
$r_c$ should be approximately half the radius at which this first 
peak occurs.
The same criterion for the largest allowed $r_c$ value was found to 
hold for other systems (not shown), viz.\ the He triplet state, the 
He isoelectronic series, and multi-electron atoms.
In a qualitative way, this condition is similar to the 
criterion used in the generation of electron-ion pseudopotentials, 
where the cut-off radius is chosen to be inside the peak of the 
valence electron density.

Also shown in Fig.~\ref{fig:HegsE} are the variational energies 
(with the true Coulomb electron-electron interaction) obtained 
for the He ground state by removing the electron cusps from the 
accurate trial wave function which was determined by variance 
minimization for use in the DMC calculations, as specified in 
Eq.~\ref{eq:smoothpsi}. 
This trial wave function includes considerable variational freedom 
in the interelectronic terms and obtains 100\% of the correlation 
energy. 
As can be seen in Fig.~\ref{fig:HegsE}, removal of the cusp out 
to radii of 0.5 a.u.\ causes little degradation of the 
quality of the energy obtained with this wave function. 
Thus the success of interelectronic terms in obtaining good 
variational energies is not primarily due to their description 
of the short-range cusp behaviour.

This trend is further borne out when we look at the convergence 
of partial CI expansions within a given single-particle basis set 
for the MCCI method.  
The CI calculations were performed with the correlation consistent 
polarized valence double-zeta (cc-pVDZ), triple-zeta (cc-pVTZ), quadruple-zeta (cc-pVQZ)
and quintuple-zeta (cc-pV5Z) basis sets of Dunning~\cite{Dunn} with 
the electron interaction $V_{\rm ps}$ for various values of $r_c$. 
All four basis sets show similar behaviour and we present only 
the cc-pVQZ results in detail.
Table~\ref{tab:heccpvqz} shows the correlation energy obtained 
as a function of cut-off radius $r_c$ for various values of the 
expansion coefficient threshold $c_{\rm min}$.
The full CI result is obtained when $c_{\rm min}=0.0$.
For larger values of the cut-off radius $r_c$ (upper rows of table),
the total correlation energy is smaller, as the variation of the 
interaction $V_{\rm ps}$ becomes smaller around the typical separation 
distance of the electrons. 
However, the fraction of the correlation energy which is recovered 
at a given CI threshold level $c_{\rm min}$ depends little on the value 
of $r_c$.

%
%

\begin{table*}

\caption{Calculated correlation energy for the He atom with a 
cc-pVQZ basis set, as a function of cut-off radius $r_c$ 
and the CI coefficient threshold $c_{\rm min}$.
The energy units are mHartree. 
The right-hand column (HF$-$QMC) shows the full correlation energy
(i.e. the difference between the Hartree-Fock energy, calculated 
with this basis, and QMC (exact) energy).
The percentages indicated are the fractions of the full 
CI correlation energy for each value of $r_c$.
}

\begin{center}
\begin{tabular}{*{8}{r@{.}l}}
\multicolumn{2}{c}{$r_c$} & \multicolumn{12}{c}{$c_{\rm min}$} & \multicolumn{2}{c}{} \\
\multicolumn{2}{c}{(a.u.)} & \multicolumn{2}{c}{$3 \times 10^{-2}$} & \multicolumn{2}{c}{$2 \times 10^{-2}$} & \multicolumn{2}{c}{$1 \times 10^{-2}$} & \multicolumn{2}{c}{$1 \times 10^{-3}$} & \multicolumn{2}{c}{$1 \times 10^{-4}$} &     \multicolumn{2}{c}{0.0}  & \multicolumn{2}{c}{HF $-$ QMC} \\
\hline
                          0&95 &        2&15  &            15&10  &             12&98  &             18&72  &             18&96  &    18&96  & \hspace*{8mm} 19&56 \hspace*{8mm}  \\
\multicolumn{2}{c}{}           &       11&01\%&            77&21\%&            66&36\% &            95&69\% &            96&94\% &    96&95\%& 100&00\% \\

                          0&80 &        6&09  &            18&45  &             16&93  &             25&11  &             25&16  &    25&16  &  25&76   \\
\multicolumn{2}{c}{}           &       23&66\%&            71&64\%&            65&74\% &            97&48\% &             97&67\%&    97&67\%& 100&00\% \\

                          0&50 &        8&64  &            27&23  &            32&79  &             36&32  &             36&32  &     36&32  &  37&18   \\
\multicolumn{2}{c}{}           &       23&24\%&            73&24\%&            88&19\% &            97&69\% &            97&69\% &    97&69\%& 100&00\% \\

                          0&20 &       14&07  &            28&93  &            35&27  &             40&75  &             40&75  &     40&75  &  41&93   \\
\multicolumn{2}{c}{}           &       33&57\%&            69&00\%&            84&13\% &            97&20\% &            97&20\% &    97&20\%& 100&00\% \\

                          0&10 &        9&37  &            28&98  &            35&35  &             40&89  &             40&89  &     40&89  &  42&18   \\
\multicolumn{2}{c}{}           &       22&22\%&            68&71\%&            83&80\% &            96&94\% &            96&94\% &    96&94\%& 100&00\% \\

                          0&00 &       27&44  &            30&61  &            36&09  &             40&90  &             40&90  &     40&90  &  42&21   \\
\multicolumn{2}{c}{}           &       65&01\%&            72&52\%&            85&50\% &            96&90\% &            96&90\% &    96&90\%& 100&00\% 
\end{tabular}
\end{center}
\label{tab:heccpvqz}
\end{table*}

\begin{figure}[tb]
\begin{center}
  \leavevmode
  \psfig{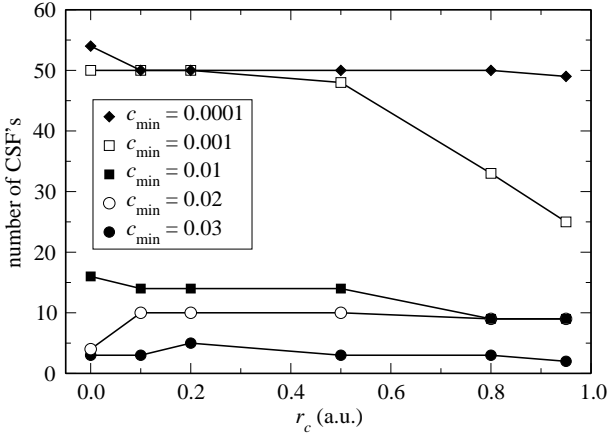}
  \vspace{0.5cm}
  \caption{
    The number of configuration state functions (CSF) in the CI 
    expansion for the He atom with the cc-pVQZ basis set as a function of $r_c$. 
    Results are presented for coefficient thresholds $c_{\rm min}$ of 
    0.03, 0.02, 0.01, 0.001 and 0.0001. 
  }
  \label{fig:Heccpvqz:csf}
\end{center}
\end{figure}

Fig.~\ref{fig:Heccpvqz:csf} shows the number of CSF's in the CI
expansion as a function of cut-off radius $r_c$ 
for various values of the expansion coefficient threshold $c_{\rm min}$. 
For values of $c_{\rm min} > 3\times 10^{-2}$, one is quickly reduced 
to a few CSF's and finally to the Hartree-Fock solution. 
As with the correlation energies, for a given threshold $c_{\rm min}$,
the number of CSF's shows relatively little change for values of 
$r_c$ between 0 and 0.95.
Again, it is worth bearing in mind that the typical electron-electron
separation is 1 a.u. and that a cut-off radius $r_c=0.95$ represents
a very smooth interaction.
Thus, removal of electron cusps does not result in a significant 
reduction in the number of contributing configurations in the CI 
expansion.

%
%

\begin{figure}[tb]
\begin{center}
  \leavevmode
  \psfig{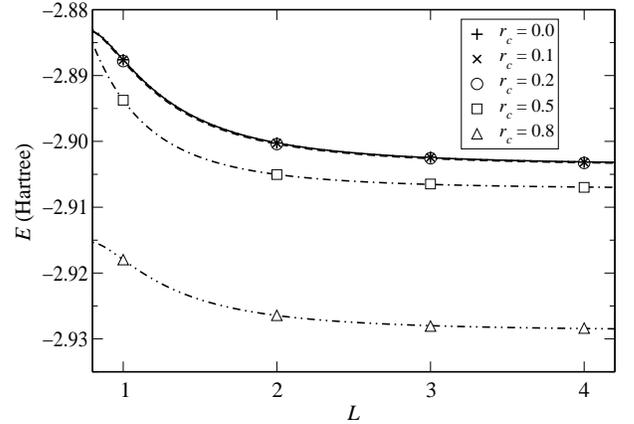}
  \vspace{0.5cm}
  \caption{
    Full CI energies $E$ for the ground state He atom as a function of $L$, the maximum
    angular momentum in the basis set used, calculated using the electron interaction 
    $V_{\rm ps}$ with values of the cut-off radius $r_c$ from 0 to 0.8.
  }
  \label{fig:ci_basis_conv}
\end{center}
\end{figure}

We next demonstrate that removal of the electron cusp does not
significantly change the convergence properties of CI expansions 
as the highest angular momentum quantum number is increased in 
the single particle basis sets. 
In Fig.~\ref{fig:ci_basis_conv}, FCI energies for He are displayed 
against increasing basis set size in the correlation consistent series 
from Dunning, and hence increasing angular momentum number in the basis. 
The convergence in the FCI energy with increasing basis set size is 
compared for electron interaction $V_{\rm ps}$, with values of the 
cut-off radius $r_c$ from 0 to 0.8.
Fitting these energies to a function similar to the asymptotic form derived in 
Ref.~\cite{Hill}, 
\begin{eqnarray}
E ~=~ E_0 + {a\over L^3}  + {b\over L^4} ~, 
\label{eq:lconv}
\end{eqnarray}
where $L$ is the maximum angular momentum in the basis,
gives best values of $a \approx 0.038$ and $b \approx -0.022$
for each value of $r_c\le0.2$ and gives 
$a \approx 0.02$ and $b \approx -0.01$ for $r_c\ge0.5$.
Thus, the predominant $1/L^3$ convergence of the energy is present for all
values of $r_c$ investigated here.

%
%

Although the asymptotic form in Eq.~\ref{eq:lconv} fits the results 
well, the parameters $a$ and $b$ do not have the asymptotic values,
$a = 0.025$ and $b = 0.008$, found for the angular momentum convergence 
associated with the short-range cusp~\cite{Morgan:CIconv}.
Two factors cause this discrepancy: (i) not only higher angular momentum 
functions are added as $L$ is increased in the Dunning basis sets;
(ii) more importantly, we are not near the asymptotic regime, where
expansion of the short-range cusp would be the dominant energy correction.

\subsection{Ground State of the Be atom}
%
%

\begin{figure}[tb]
\begin{center}
  \leavevmode
  \psfig{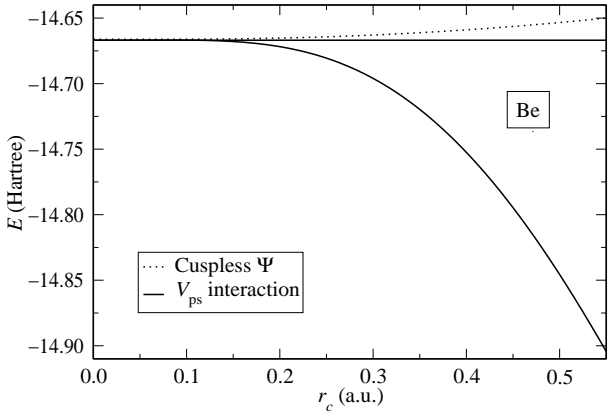}
  \vspace{0.5cm}
  \caption{
    Be atom ground state energy $E$ as a function of 
    cut-off radius $r_c$, obtained from (a) DMC calculations 
    using the norm-conserving pseudo-interaction $V_{\rm ps}$ (solid line);
    (b) variational calculations using the true Coulomb interaction 
    upon removal of the electron cusp from the trial wave function 
    (see Appendix~\protect\ref{app:srjast}) used in the DMC calculation (dashed line).
  } 
  \label{fig:BegsE}
\end{center}
\end{figure}

The qualitative nature of the CI energy convergence seen in the
previous section is not special to the He atom.
Fig.~\ref{fig:BegsE} shows the DMC calculated ground state energies 
of Be for various cut-off radii $r_c$. 
Because of the much smaller radius of the 1s orbital in Be, compared
to He, the energetically appropriate cut-off radius $r_c$ for the 
pseudo-interaction is substantially smaller.
Also shown in Fig.~\ref{fig:BegsE} are the variational energies 
obtained for the Be ground state by removing the electron cusps 
(as defined in Eq.~\ref{eq:smoothpsi}) from the accurate trial 
wave function which was determined by variance minimization for use 
in the DMC calculations. 
This trial wave function includes interelectronic terms and 
obtains 99.5\% of the correlation energy. 
Removal of the cusp causes significant degradation of the quality 
of the wave function only for $r_c>0.3$ atomic units, which is 
comparable to the typical separation of the 1s electrons.
Thus, in Be, as in He, the success of interelectronic terms in 
obtaining good variational energies is not primarily due to their 
description of the short-range cusp behaviour. 

%
%

CI calculations using basis sets with s- and p-functions 
(the TZP basis of Ahlrichs {\it et al.}\cite{tmole_basis}), 
s-, p- and d-functions (6-311g* \cite{6311g}), 
s-, p-, d- and f-functions
(the ANO basis of Widmark {\it et al.} \cite{be_ano}) 
and s-, p-, d-, f- and g-functions 
(i.e. the ANO basis, augmented with a g-function with exponent 1.2) 
were carried out. 
The values of the CI coefficient threshold $c_{\rm min}$ were taken 
to be $10^{-2}$, $10^{-3}$ and $10^{-4}$ for all basis sets. 
Calculations were carried out at cut-off radii 
$r_c$ of 0 (i.e. Coulomb interaction), 0.1 and 0.2 a.u.

\begin{figure}[tb]
\begin{center}
  \leavevmode
  \psfig{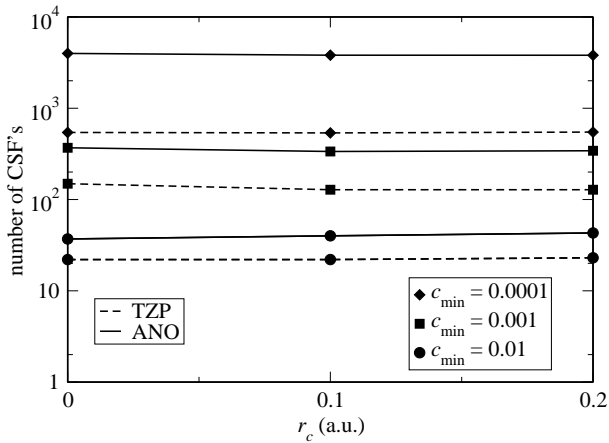}
  \vspace{0.5cm}
  \caption{
    The number of configuration state functions (CSF) in the CI 
    expansion for the Be atom with TZP and ANO basis sets. 
    Results are presented for coefficient thresholds 
    $c_{\rm min}$ of 0.01, 0.001 and 0.0001 
    at cut-off radii $r_c$  of 0.0, 0.1 and 0.2 a.u.
  }
  \label{fig:Be:csf}
\end{center}
\end{figure}

In Tables~\ref{tab:Betzp} and \ref{tab:Beano} we present the 
correlation energy for Be, obtained with the TZP and ANO basis 
sets, respectively.
Fig.~\ref{fig:Be:csf} shows the 
number of CSF's as a function of the cut-off radius $r_c$ 
for various values of the CI coefficient threshold $c_{\rm min}$ 
in the MCCI approach.
We see that, especially at the higher levels of accuracy 
(i.e. small $c_{\rm min}$), neither the fraction of the correlation 
energy obtained nor the number of CSF's used depend strongly on 
the cut-off radius $r_c$.
As we saw in the He atom, removing the divergence of the
Coulomb electron interaction does not substantially change
the CI energy convergence at this level of accuracy.

%
%

\begin{table}[t]

\caption{
Calculated correlation energy, as a function of cut-off radius 
$r_c$ and CI coefficient threshold $c_{\rm min}$, for the Be atom 
with the TZP basis set.
All energies are in mHartree.
The right-hand column shows the QMC correlation energy 
(i.e. the difference between the Hartree-Fock energy for this 
basis and the QMC energy) for each value of the cut-off radius 
$r_c$.
The percentages indicated are the fractions of the QMC 
correlation energy for each value of $r_c$.
}

\begin{center}
\begin{tabular}{*{5}{r@{.}l}}
\multicolumn{2}{c}{$r_c$} & \multicolumn{6}{c}{$c_{\rm min}$} & \multicolumn{2}{c}{} \\
\multicolumn{2}{c}{(a.u.)} & \multicolumn{2}{c}{$1 \times 10^{-2}$} & \multicolumn{2}{c}{$1 \times 10^{-3}$} & \multicolumn{2}{c}{$1 \times 10^{-4}$} & \multicolumn{2}{c}{HF $-$ QMC} \\
\hline
                  0&20   &             51&63  &             60&91  &              61&61  &  \hspace*{8mm}  99&39 \hspace*{8mm} \\
\multicolumn{2}{c}{}     &           51&94\%  &           61&28\%  &            61&99\%  & 100&00\%  \\

                  0&10   &             51&55  &             60&81  &              61&53  &    95&01  \\
\multicolumn{2}{c}{}     &           54&26\%  &           64&01\%  &            64&76\%  & 100&00\%  \\

                  0&00   &             56&42  &             61&51  &              62&01  &    94&89  \\
\multicolumn{2}{c}{}     &           59&46\%  &           64&82\%  &            65&35\%  & 100&00\%  \\
\end{tabular}
\end{center}

\label{tab:Betzp}
\end{table}

%
%

\begin{table}[t]

\caption{
Calculated correlation energy for the Be atom with the ANO basis set.
The notation is the same as in Table~\ref{tab:Betzp}.
}

\begin{center}
\begin{tabular}{*{5}{r@{.}l}}
\multicolumn{2}{c}{$r_c$} & \multicolumn{6}{c}{$c_{\rm min}$} & \multicolumn{2}{c}{} \\
\multicolumn{2}{c}{(a.u.)} & \multicolumn{2}{c}{$1 \times 10^{-2}$} & \multicolumn{2}{c}{$1 \times 10^{-3}$} & \multicolumn{2}{c}{$1 \times 10^{-4}$} & \multicolumn{2}{c}{HF $-$ QMC} \\
\hline
                  0&20   &             53&40  &             71&35  &              74&13  &  \hspace*{8mm}  90&39 \hspace*{8mm} \\
\multicolumn{2}{c}{}     &           59&07\%  &           78&93\%  &            82&01\%  & 100&00\%  \\

                  0&10   &             42&54  &             71&97  &              74&80  &    94&43  \\
\multicolumn{2}{c}{}     &           45&05\%  &           76&22\%  &            79&21\%  & 100&00\%  \\

                  0&00   &             58&46  &             72&30  &              74&66  &    93&91  \\
\multicolumn{2}{c}{}     &           62&25\%  &           76&99\%  &            79&50\%  & 100&00\%  \\ 
\end{tabular}
\end{center}

\label{tab:Beano}
\end{table}

%
%

\section{Conclusions}

In this paper we have directly explored the effect of the electron 
cusp on the convergence of the energy for CI wave functions.
To do so, we introduced a fictitious electron-electron interaction 
which, unlike the true Coulomb potential, does not diverge at 
electron coalescences and therefore has many-body eigenfunctions 
which are smooth there.
The smooth potential is obtained by a method analogous to the 
well-established treatment of electron-ion interactions with smooth, 
norm-conserving pseudopotentials. 
These give scattering properties
very similar to the true potential and improve the convergence 
of plane-wave expansions for single-particle wave functions.
By analogy, one might expect that a similar treatment of the
electron-electron interaction would improve the convergence
of CI expansions.

For the He and Be atoms, we have compared the CI expansions of 
wave functions for the true, Coulomb interaction and for smooth
pseudo-interactions between the electrons.
We have also compared the variational energies of highly accurate
wave functions containing $r_{12}$ terms, where the cusp condition 
is exactly satisfied, with the energy for the same wave functions, 
from which the cusp has been deliberately removed out to a given radius.

The main result of our study is that a description of the electron 
cusp as such is {\it not} a limiting factor in calculating correlation 
effects with configuration interaction methods at the mHartree 
level of accuracy.
We are led to this conclusion since the replacement of the divergent 
Coulomb interaction with a finite interaction leaves the convergence 
properties largely unchanged. 
The slow convergence of CI expansions in this energy range must be 
attributed to medium-range correlations, which are present for both
types of electron interaction. 
The results obtained by explicitly removing the cusp from 
accurate variational wave functions support this conclusion,
showing that the cusp of the wave function is not energetically 
important. 
This is found, even when cusp-like behaviour is removed out to 
a distance of half the typical interelectronic separation.

When describing correlation effects with explicit $r_{12}$ methods, 
it is a mistaken notion to assign the improvement in convergence to 
their ability to describe the short-range cusp. 
Rather, the improvement in convergence at the mHartree level
must be understood in terms of a better description of medium-range 
correlations when expressed in interelectronic coordinates. 
This fact must be considered when developing theoretical treatments 
of the electronic correlation problem. 

\section{Acknowledgments}

We wish to thank D.~R.~Hamann for stimulating discussions.
This work has been supported by Enterprise Ireland Grant Number
SC/98/748 and by the Irish Higher Education Authority.

%

%
%
\appendix
%
%
\section{Generation of Electron Pseudo-interactions}
\label{app:gen_psint}
%
%
%

The procedure for constructing the electron pseudo-interaction 
$V_{\rm ps}$ follows Hamann's construction~\cite{Hamann:gncpp}
of generalised norm-conserving pseudopotentials for
electron-ion scattering, with two modifications: 
(i) Only two particles are involved in the generation of the electron-electron
pseudo-interaction, so that we do not need to include screening potentials;
(ii) because we are solving the scattering problem in the centre of
mass frame of the electrons and use the relative coordinate ${\bf r}$ 
of the electrons, the mass of the electron is replaced by the reduced 
mass for the two-electron system.
In Hartree atomic units, the interaction potential is 
$V(r) = 1/r$ and the reduced mass is $1/2$. 
The radial Schr\"{o}dinger equation for the relative motion of two
electrons is then
\begin{eqnarray}
 -\frac{d^2u_l}{dr^2} + \left(\frac{l(l+1)}{r^2}+(V-\epsilon_l)\right)u_l=0 ,
 \label{eq:rSchr}
\end{eqnarray}
where $u_l(r)$ is $r$ times the wave function in the relative coordinates,
with angular momentum $l$ and energy $\epsilon_l$ (the kinetic 
energy of the separated electrons in the centre of mass frame).

In the generation of a pseudopotential, the true potential and
pseudopotential become identical outside a suitably chosen cut-off 
radius $r_c$ and are constructed to have identical scattering phase 
shifts and energy derivative of the phase shift at a reference 
energy $\epsilon_l$.
A smooth pseudo wave function $u_{\rm ps}(r)$ is generated numerically
to ensure ``norm-conservation'', 
i.e. $u_{\rm ps}(R_\infty) = u_l(R_\infty)$ and
\begin{eqnarray}
\int_0^{R_\infty} u_{\rm ps}(r)^2 dr ~= ~\int_0^{R_\infty} u_l(r)^2 dr ~,
 \label{eq:norm}
\end{eqnarray}
for all $R_\infty >> r_c$.
In practice, we choose $R_\infty \sim 2.5 r_c$, as in 
Ref.~\cite{Hamann:gncpp}.
Given $u_{\rm ps}$, a smooth, finite pseudopotential $V_{\rm ps}$ 
is generated by inversion of the Schr\"odinger Eq.~\ref{eq:rSchr} 
for $r<R_\infty$, with $u_l$ and $V$ substituted by $u_{\rm ps}$ 
and $V_{\rm ps}$, respectively.
We follow the steps exactly as given in Sec.~II of 
Ref.~\cite{Hamann:gncpp}, except where the reduced mass enters
in the final inversion, slightly modifying Eq.~12 of Hamann's work. 

In the norm-conserving pseudopotential approach, different
scattering potentials are often generated for different values of
the angular momentum $l$.
The full many-body wave function is anti-symmetric with respect to
exchange of any two electrons. 
Thus, if the two electrons have parallel spin, the orbital part of 
the wave function must have odd parity and the allowed angular 
momentum $l$ must be odd.
Thus, the dominant parallel-spin scattering has $l=1$.
For singlet (anti-parallel) scattering, the orbital part is 
symmetric and $l$ must be even, with dominant allowed angular momentum 
$l=0$.
Following the usual pseudopotential approach, we might then expect 
to generate a parallel-spin scattering potential using $l=1$ and an
anti-parallel-spin scattering potential using $l=0$. 
In practice, we find that the $l=0$ potential gives good scattering 
for both angular momenta and the use of separate parallel- and 
anti-parallel-spin potentials is unnecessary.

Since $\epsilon_l$ is positive, the wave function will oscillate
for large $r$, and if a node occurs for $r<R_{\infty}$, 
the inversion of the Schr\"{o}dinger equation will fail. 
Thus, the joint requirements,
that $r_c$ be sufficiently large to produce a smooth 
pseudo-interaction and that $R_{\infty} \stackrel{>}{\sim} 2.5 r_c$,
effectively limit the maximum reference energy $\epsilon_l$ at 
which the pseudo-interaction construction is possible.

We find that the pseudo-interaction is transferable (i.e.\ has very 
similar scattering strength to the true interaction) for energies over 
a wide range about the reference energy $\epsilon_l$, as found 
previously for electron-ion pseudopotentials.
In particular, we find that transferability is best for energies less
than the reference energy at which the pseudopotential was generated.
This would suggest that the best reference energy to use is the largest
energy compatible with the procedure for inversion of the radial 
Schr\"odinger equation, a choice confirmed by calculations of many-body
eigenstates using pseudo-interaction potentials generated with different 
values of the reference energy.

%
%
\section{First order perturbation analysis of pseudo-interactions}
\label{app:acc_psint}
%
%

In this Appendix, we consider the first-order error in the energy expected for 
a typical, arbitrary smooth electron-electron potential which is equal 
to the exact interaction $1/r$ outside some (adjustable) cut-off radius $r_c$.
We will assume that the potential is (at least approximately) of the form 
\begin{eqnarray}
 V_{r_c}(r) ~=~ {1\over r_c} U(r/r_c) 
\label{eq:psform}
\end{eqnarray}
where $U(r/r_c)$ is some universal function independent of $r_c$.
Clearly both the true interaction $1/r$ and the smooth interaction $V_{\rm erf}$
are of this form exactly.
$V_{\rm ps}$ is also approximately of this form.
Thus, the difference $\Delta V$ between the interaction $V_{r_c}(r)$ and the
true electron-electron interaction $1/r$ is of the same form:
\begin{eqnarray}
 \Delta V(r) ~=~ {1\over r_c} \Delta U(r/r_c) ,
\label{eq:psdiff}
\end{eqnarray}
where $\Delta U(\rho)$ is zero for $\rho>1$.

Replacing the true electron-electron interaction with $V_{r_c}(r)$ then
gives rise to a first-order energy error:
\begin{eqnarray}
\delta\epsilon^{(1)} & = & \langle \Psi |\Delta V| \Psi \rangle  \nonumber \\
 & = & \langle\Psi|\left(V_{r_c} - \frac{1}{r}\right)|\Psi\rangle,
\label{eq:firstord}
\end{eqnarray}
where $|\Psi\rangle$ is the eigenstate of the many-electron system.
If $P(r)$ is the electron pair distribution function (averaged at
the pair separation distance $r$), the energy error may be rewritten as:
\begin{eqnarray}
\delta\epsilon^{(1)} & = & 
\int_{0}^{\infty} P(r) {1\over r_c} \Delta U(r/r_c) 4\pi r^2 dr \nonumber \\
                     & = & 
r^2_c \int_{0}^{1} P(r_c\rho) \Delta U(\rho) 4\pi \rho^2 d\rho .
\label{eq:firstordn}
\end{eqnarray}
If $r_c$ is sufficiently small and the pair-distribution function $P(r)$
does not tend to zero at $r=0$, then $P(r_c\rho)$ may be replaced by $P(0)$
in the integral and $ \delta\epsilon^{(1)}  \approx K P(0) r_c^2 $ for small
$r_c$, where $K$ is a constant for the type of interaction, defined by
\begin{eqnarray}
K    ~\equiv~    \int_{0}^{1} \Delta U(\rho) 4\pi \rho^2 d\rho .
\label{eq:Kdef}
\end{eqnarray}
The construction of $V_{\rm ps}$ ensures that the energy error is
identically zero for the two-electron scattering at the reference
energy, and therefore $K = 0$ for $V_{\rm ps}$.
However, for an arbitrary smooth interaction (such as $V_{\rm erf}$), 
$K$ will not be zero.

%
%
\section{Short-range correlation term obtained from norm-conserving potential}
\label{app:srjast}
%
%

In the generation of the norm-conserving potential discussed in
Appendix~\ref{app:gen_psint}, we obtained a pseudo wave function 
$u_{\rm ps}(r)$, which is a solution to the radial Schr\"odinger 
equation at the scattering energy $\epsilon_l$ for the pseudo-interaction, 
and the wave function $u_l(r)$, which is the corresponding solution 
for the true Coulomb interaction.
The ratio $J_{sr} = u_l/u_{\rm ps}$ of these two functions has the correct 
electron cusp (i.e. logarithmic derivative of $-1/4$ or $-1/2$ for
parallel or anti-parallel spins, respectively, at $r=0$) and tends
rapidly to $1$ for $r>r_c$. 
This function $J_{sr}$ may be said to contain the short-range cusp 
behaviour of the Coulomb scattering solution, since it represents the 
difference between the scattering wave function for the divergent 
potential and that for the finite pseudo-interaction.

Moreover, if $\Psi({\bf r}_1, \dots , {\bf r}_N)$ is a many-electron
wave function (e.g.\ one which has inter-electronic coordinates $r_{ij}$
explicitly included) with the correct electron cusps, the function 
\begin{eqnarray}
\Psi_s = \Psi({\bf r}_1, \dots , {\bf r}_N)/\prod_{i<j} J_{sr}(r_{ij}) 
\label{eq:smoothpsi}
\end{eqnarray}
is smooth at electron coalescences but has very similar electron
correlations for $r_{ij}>r_c$.
The wave function $\Psi_s$ has the electron cusp behaviour 
of the original wave function $\Psi$ removed inside the cut-off 
radius $r_c$, but is otherwise the same as $\Psi$.
Thus, the energy difference 
$\langle\Psi_s|H|\Psi_s\rangle -  \langle\Psi|H|\Psi\rangle $
is a good measure of the energy cost of removing the electron
cusp behaviour for $r<r_c$ from a correlated wave function, while
keeping other aspects of the wave function (e.g. any benefits there 
may be to the inclusion of inter-electronic coordinates in the form 
of the trial function) unchanged.

%
%
\section{Evaluation of Two-Electron Gaussian Integrals with Pseudo-Interactions}
\label{app:two_cent_int}
%
%

In this Appendix, we give details of the evaluation of two-electron 
Gaussian basis set integrals for the electron interactions, 
$V_{\rm erf}$ and $V_{\rm ps}$, and their incorporation into existing 
electron integral programs.

For the interaction $V_{\rm erf}$, the necessary modification of 
two-electron integrals for all s-type basis functions have been given 
by Savin~\cite{Savin:erf}.
We do not use Savin's approach, but rather directly modify the
Rys quadrature approach of Dupuis, Rys, and King~\cite{drk},
to obtain matrix elements for the interaction $V_{\rm erf}$
in place of those for $1/r_{12}$. 
To evaluate two-electron integrals for the pseudo-interaction 
$V_{\rm ps}$, we use the form given in Eq.~\ref{eq:psfit}. 
This involves the evaluation of integrals with the interaction 
$V_{\rm erf}$ and a sum of Gaussian terms.
In codes (e.g. the ARGOS code~\cite{COLUMBUS}) which use the Rys 
quadrature~\cite{drk} approach for the evaluation of two-electron 
integrals with the Coulomb interaction, 
these Gaussian terms are already evaluated in an intermediate stage
and may be used in the present context.

Let us define two-electron integrals of the Coulomb interaction as 
\begin{eqnarray*}
\lefteqn{(\phi_i \phi_j |\frac{1}{r_{12}}| \phi_k \phi_l)} \nonumber \\
& \equiv & 
\int \!\! \int \phi_i (r_1) \phi_j (r_1) \frac{1}{r_{12}} \phi_k (r_2) \phi_l (r_2) dr_1 dr_2 ~,
\end{eqnarray*}
for Gaussian basis functions, 
$$\phi_i(r) = {\rm exp}(-a_i |r-R_i|^2) \prod_{x,y,z}(x-X_i)^{n_{xi}}~.$$ 
The Rys scheme represents the $1/r_{12}$ term in Gaussian integral form, 
to obtain: 
\begin{equation}
(\phi_i\phi_j| \frac{1}{r_{12}} |\phi_k\phi_l) = 
\frac{2}{\sqrt \pi} \int_0^\infty du 
(\phi_i\phi_j| exp(-u^2 r_{12}^2) |\phi_k\phi_l) .
\label{eq_2einti}
\end{equation}
This integral is then transformed to an integral over the variable 
$t$, where $t^2 = u^2/(\rho+u^2)$ and 
$1/\rho = 1/(a_i + a_j) + 1/(a_k + a_l)$:
\begin{equation}
(\phi_i\phi_j| \frac{1}{r_{12}} |\phi_k\phi_l) = 
 \int_0^1  P_L(t) \exp(-Xt^2) \; dt ~,
\label{eq_2eintt}
\end{equation}
where $P_L$ is a polynomial and $X = \rho|r_A - r_B|^2$.
Here, $r_A$ and $r_B$ are the usual weighted-average centers
of the Gaussian orbital pairs $(i,j)$ and $(k,l)$, respectively,
as given in Ref.~\cite{drk}.
The integration over $t$ is then evaluated exactly by an 
$n$-point quadrature formula:
\begin{eqnarray*}
(\phi_i\phi_j| \frac{1}{r_{12}} |\phi_k\phi_l) = 
 \sum_{\alpha=1}^n  P_L(t_\alpha) W_\alpha  ~,
\end{eqnarray*}
where $t_\alpha(X)$ and $W_\alpha(X)$ are determined as in
Ref.~\cite{drk}.

We observe that the interaction $V_{\rm erf}$ may be written in 
exactly the same form as in Eq.~\ref{eq_2einti}, replacing the upper 
limit of integration ($u=\infty$) with $u=1/r_c$.
This changes the upper limit of integration in the variable $t$ 
from 1 to $t_c = 1/\sqrt{1+\rho r_c^2}$ in Eq.~\ref{eq_2eintt}.
To use directly the Rys quadrature approach, we further transform to 
the variable $t' = t/t_c$, to obtain
\begin{eqnarray*}
(\phi_i\phi_j| \frac{{\rm erf}(r_{12}/r_c)}{r_{12}} |\phi_k\phi_l) = 
 t_c \int_0^1  P_L(t_ct') \exp(-Xt_c^2t'^2) \; dt' ~,
\end{eqnarray*}
which can be evaluated as,
\begin{eqnarray*}
(\phi_i\phi_j| \frac{{\rm erf}(r_{12}/r_c)}{r_{12}} |\phi_k\phi_l) = 
 \sum_{\alpha=1}^n  P_L(t'_\alpha) W'_\alpha ~,
\end{eqnarray*}
where the quadrature points $t'_\alpha$ and weights $W'_\alpha$
are found by defining $X' = Xt_c^2$, $t'_\alpha = t_c t_\alpha(X')$,
and $W'_\alpha = t_c W_\alpha(X')$.

We can similarly evaluate the Gaussian terms in the
representation of $V_{\rm ps}$. 
Defining the integrand in Eq.~\ref{eq_2einti}, 
${\cal F}(u) \equiv (\phi_i\phi_j| exp(-u^2 r_{12}^2) |\phi_k\phi_l)$,
we see that $ \sum P_L(t_\alpha) W_\alpha = \sum A_\alpha {\cal F}(u_\alpha)$,
where $u_\alpha^2/\rho = t_\alpha^2/(1-t_\alpha^2)$ and 
$A_\alpha = (2/\sqrt\pi) [(\rho+u^2)^{3/2} / \rho] \exp[Xu^2/(\rho+u^2)]W_\alpha(X)$. 
One may directly modify this finite-point quadrature summation to 
evaluate two-electron integrals for the Gaussian terms, 
replacing $n$ with $n_G$, $A_\alpha$ with $c_i$, and $u_\alpha^2$ with $d_i$.  
We have applied these modifications to the two-electron integrals 
within the ARGOS code, part of the COLUMBUS program package~\cite{COLUMBUS}.

%


\begin{thebibliography}{99}
%

\bibitem{KlahnMorg} B. Klahn and J.D. Morgan III, 
J. Chem. Phys. {\bf 81}, 410 (1984).

\bibitem{Hill} R. N. Hill, J. Chem. Phys. {\bf 83}, 1173 (1985).

\bibitem{Morgan:CIconv} J. D. Morgan, in 
{\it Numerical Determination of the Electronic Structure of Atoms, 
Diatomic, and Polyatomic Molecules}, edited by M. Defranceschi, and 
J. Delhalle (Kluwer Academic Publishers, Dordrecht, 1989) p. 49.

\bibitem{r12} W. Kutzelnigg and W. Klopper,  J. Chem. Phys. {\bf 94}, 1985 (1991).

\bibitem{KutzMorg} W. Kutzelnigg and J. D. Morgan III, 
J. Chem. Phys. {\bf 96}, 4484, (1992).

\bibitem{Hoff} M. Hoffmann-Ostenhoff, T. Hoffman-Ostenhof, and H. Stremnitzer,
Phys. Rev. Lett. {\bf 68}, 3857 (1992).

\bibitem{Helgaker} T. Helgaker, W. Klopper, W. Koch, and J. Noga,
J. Chem. Phys. {\bf 106}, 9639 (1997).

\bibitem{Lein} T. Leininger, H. Stoll, H-J. Werner, and A. Savin, 
Chem. Phys. Lett. {\bf 275}, 151 (1997).

\bibitem{NooBar} M. Nooijen and R. J. Bartlett, 
J. Chem. Phys. {\bf 109}, 8232 (1998).

\bibitem{Silanes} I. Silanes, J. M. Ugalde, and R. J. Boyd, 
J. Mol. Struct. (THEOCHEM), {\bf 527} 27 (2000).

\bibitem{Kato:cusp} T. Kato, Commun. Pure Applied Math. {\bf 10}, 151 (1957).

\bibitem{Hyll}E. A. Hylleraas, Z. Phys. {\bf 51}, 347 (1929).

\bibitem{gem_gauss} R. Bukowski, B. Jeziorski, and K. Szalewiczi, 
J. Chem. Phys. {\bf 100}, 1366, (1994).

\bibitem{BFHM90} J. D. Baker, D. E. Freund, R. N. Hill, and J. D. Morgan III, 
Phys. Rev. A {\bf 41}, 1247 (1990).

\bibitem{Gil63} T. L. Gilbert, Rev. Mod. Phys. {\bf 35}, 491 (1963).

\bibitem{King} H. F. King, Theor. Chim. Acta {\bf 94}, 345 (1996).

\bibitem{HSC} D. R. Hamann, M. Schl\"uter, and C. Chiang,
Phys. Rev. Lett. {\bf 43}, 1494 (1979).

\bibitem{Pickett} W. E. Pickett, Comp. Phys. Rep. {\bf 9}, 115 (1989).

\bibitem{Hamann:gncpp} D. R. Hamann, Phys. Rev. B {\bf 40}, 2980 (1989).

\bibitem{Savin:erf} A. Savin, in 
{\it Recent Developments and Applications of Modern Density Functional Theory} 
edited by J. M. Seminario (Elsevier, 1996) p. 327.

\bibitem{Fil-Umr} C. Filippi and C. J. Umrigar, 
J. Chem. Phys. {\bf 105}, 213 (1996).

\bibitem{Cyrus:varmin} C. J. Umrigar, K. G. Wilson, and J. W. Wilkins, 
Phys. Rev. Lett. {\bf 60}, 1719 (1987).

\bibitem{Gre1} J. C. Greer, J. Chem. Phys. {\bf 103}, 1821 (1995).

\bibitem{Gre2} J. C. Greer, J. Comp. Phys. {\bf 181}, 146 (1998).

\bibitem{drk} M. Dupuis, J. Rys, and H. F. King, 
J. Chem. Phys. {\bf 65}, 111 (1976).

\bibitem{Davidson} E. R. Davidson, S. A. Hagstrom, S. J. Chakrvorty,
V. M. Umar, and C. F. Fischer, Phys. Rev. A {\bf 44}, 7071 (1991).

\bibitem{Dunn} T. H. Dunning Jr., J. Chem. Phys. {\bf 90}, 1007 (1989)

\bibitem{tmole_basis} A. Schaefer, H. Horn, and R. Ahlrichs, 
J. Chem. Phys. {\bf 97}, 2571 (1992).

\bibitem{6311g}  R. Krishnan, S. J. Binkley, R. Seeger, and J. A. Pople,  
J. Chem. Phys. {\bf 72}, 650 (1980).

\bibitem{be_ano} P. O. Widmark, P. A. Malmqvist, and B. O. Roos, 
Theor. Chim. Acta {\bf 77}, 291 (1990).

\bibitem{COLUMBUS} R. Shepard, I. Shavitt, R. M. Pitzer, D. C. Comeau, 
M. Pepper, H. Lischka, P. G. Szalay, R. Ahlrichs, F. B. Brown, and J. Zhao,
Int. J. Quant. Chem. Symp.  {\bf 22}, 149 (1988).




\end{thebibliography}
\end{document}